\documentclass{PoS}

\title{The Synergy between VLBI and Gaia astrometry}

\ShortTitle{The Synergy between VLBI and Gaia astrometry}

\author{\speaker{Huib van Langevelde}$^{ab}$, Luis Henry Quiroga$-$Nu\~nez$^{ba}$
Wouter Vlemmings$^c$,
Laurent Loinard$^d$,
Mareki Honma$^e$,
Akiharu Nakagawa$^f$,
Katharina Immer$^a$,
Ross Burns$^{agh}$,
Ylva Pihlstr\"om$^i$,
Lorant Sjouwerman$^j$,
R.~Michael Rich$^k$,
Iniyan Natarajan$^{l}$, 
Roger Deane$^{lm}$\\
\llap{$^a$}JIVE, Oude Hoogeveensedijk 4, 7991 PD Dwingeloo, the Netherlands\\
\llap{$^b$}Sterrewacht Leiden, Leiden University, Postbus 2300, 9513 RA Leiden, the Netherlands\\
\llap{$^c$}Dept.\ of SEE, Chalmers Univ., Onsala Space Observatory, SE-439 92 Onsala,
Sweden\\
\llap{$^d$} Instituto de Radioastronom\'ia y Astrof\'isica, Universidad Nacional Aut\'onoma de M\'exico,  58089 Morelia, Michoac\'an, M\'exico\\
\llap{$^e$}Mizusawa VLBI Obs., NAOJ, 2-12 Hoshigaoka-cho, Mizusawa, Oshu, Iwate 023-0861, Japan\\
\llap{$^f$}Graduate School of Science and Engineering, Kagoshima University, 1-21-35 Korimoto, Kagoshima-shi, Kagoshima 890-0065, Japan\\
\llap{$^g$}Mizusawa VLBI Observatory, National Astronomical Observatory of Japan, 2-21-1 Osawa, Mitaka, Tokyo 181-8588, Japan\\
\llap{$^h$}Korea Astronomy and Space Science Institute 776, Daedeokdae-ro, Yuseong-gu, Daejeon, 34055, Republic of Korea\\
\llap{$^i$}Dept.\ of Phys. \& Astronomy, Univ.\ of New Mexico, MSC07 4220, Albuquerque, NM 87131 USA\\
\llap{$^j$}NRAO, P.O.~Box 0, Lopezville Rd 1003, Socorro, NM 87801 USA\\
\llap{$^k$}Dept.\ of Phys. \& Astronomy, Univ.\ of California, Los Angeles, CA 90095-1547 USA\\
\llap{$^l$}CfRAT\&T, Dept. of Phys. \& Elect., Rhodes University,-
Grahamstown 6140, South Africa\\
\llap{$^m$}Department of Physics, University of Pretoria, Hatfield, Pretoria, 0028, South Africa\\
E-mail: \email{langevelde@jive.eu}}

\abstract{With the publication of Gaia DR2, 1.3 billion stars now have public parallax and proper motion measurements. In this contribution, we compare the results for sources that have both optical and radio measurements, focusing on circumstellar masers. For these large, variable and bright AGB stars, the VLBI astrometry results can be more robust. Moreover, there are a number of applications where VLBI astrometry provides unique data for studying stellar populations and Galactic structure. The BeSSel project not only provides parallax and proper motions at much larger distances than Gaia can reach, but it also uniquely samples the spiral arms of the Galaxy. The evolved stars in the BAaDE sample can potentially constrain the dynamics and stellar content of the inner bulge and bar of the Milky Way, not reachable in the optical.}

\FullConference{14th European VLBI Network Symposium \& Users Meeting (EVN 2018)\\
		8-11 October 2018\\
		Granada, Spain}

\begin{document}

\section{Introduction}

Accurate, milli-arcsecond astrometry at radio wavelengths is key for addressing a number of astrophysical problems. Recent contributions with the EVN range from lining up a radio detection of FRBs with a distant dwarf galaxies\cite{marcotefrb} and gravitational wave NS-NS mergers (e.g.~\cite{ghirlandansns}), to the dynamics of High Mass Star Formation (HMSF) regions (e.g. \cite{goddiw3}). At the same time, by providing $20-50$ mirco-arcsecond accurate positions for over a billion stars, the Gaia mission is revolutionising astrometry in the optical. It is providing parallaxes and proper motions for extremely large samples of stars, complemented with photometry and variability data. Not only has this resulted in new calibrations of the properties of specific stellar populations, but also in novel insights into the distribution of these stellar populations across the Galaxy (e.g. \cite{blandhawthorn,helmi}). These results contribute in a fundamental way to our understanding of how the Milky Way was built up over cosmological ages. Disentangling its merger history is an exciting way to observe the structure formation in the Universe.
But even with this (DR2) deluge of Gaia results, doing stellar astrometry with VLBI remains of fundamental interest and has important synergy with these recent optical astrometry studies. Although doing astrometry with VLBI is very elaborate compared to querying the Gaia databases, we note that the best VLBI measurements are still more accurate than those obtained by Gaia. 

\begin{figure}[b]
\centering
\includegraphics[width=.75\textwidth]{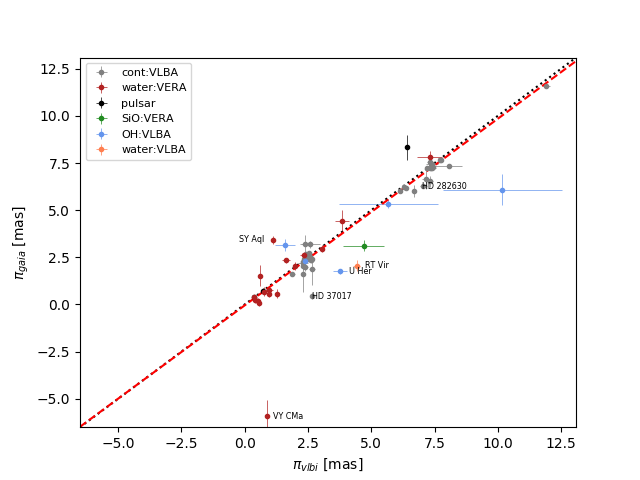}
\caption{A direct comparison of parallaxes obtained in the radio with VLBI techniques and in the optical as published in the Gaia DR2. Continuum sources are pre-main sequence sources from \cite{gob1,gob2,gob3}, VERA results for water and SiO results from \cite{kamezaki12, kamezaki16, nakagawa08, nakagawa14, nakagawa16, nyu11, min14}, VLBA OH masers from \cite{vlemmings04,vlemmingsrefined}, and VLBA water masers from \cite{zhang17, kuruyama05}, pulsar results from references in \cite{jennings18}}
\label{fig1}
\end{figure}

Admittedly, VLBI can only be done on sources that harbour bright, non-thermal radio emission. This implies that additional assumptions must be made about the origin of the radio synchrotron or molecular maser emission that is used as a beacon, where Gaia usually (but not always!) directly measures the position of the mass centre of a star. Nevertheless, the most important reason for doing astrometry with VLBI is that it is unaffected by interstellar (or circumstellar) extinction. It can therefore probe the inner Galaxy, its bar and bulge, at Galactic latitudes where Gaia's range is restricted to $\le$4 kpc from the Sun. Moreover, water and methanol masers are associated with embedded HMSF sites, providing an opportunity to probe dark Giant Molecular Clouds (GMC) and thereby the rotation of the Galaxy and the location of its spiral arms.

\begin{figure}[t]
\centering
\includegraphics[width=.6\textwidth]{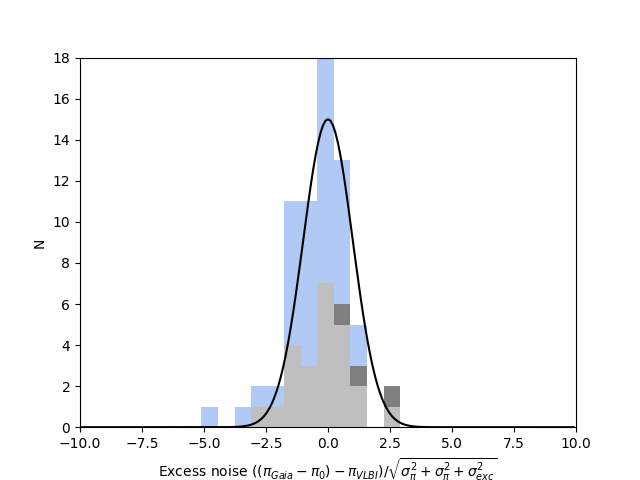}
\caption{The residuals for Gaia versus VLBI parallaxes, allowing for a Gaia parallax zero point of $-46\mu$as. The distribution has been normalised by the quadratically summed errors, including the Gaia estimate of "excess noise". In blue are AGB stars, grey radio continuum from pre-main sequence stars, black are binary pulsars (see Fig.~1 for the data references).}
\label{fig2}
\end{figure}

\section{VLBI-Gaia comparison}

For a number of objects the Gaia DR2 allows a direct comparison with VLBI parallax and proper motion results \cite{gaiadr2}. This includes a number of active stars with non-thermal radio continuum, for example active pre-main sequence stars (Fig.~1). Also, some pulsars have optical companions. The analysis presented here, however, focuses on evolved stars with circumstellar masers. Because the masers occur {---}depending on molecular species{---} at considerable distance from the optically visible star, the proper motions will generally not be exactly the same, except when this is a stationary offset. However, maser components with a linear motion with respect to the star will still show the same parallax. In Fig.~1, we show a plot of VLBI parallax versus Gaia parallax for a number of evolved stars combined with a number of pre-main sequence stars and binary pulsars for comparison. In general it is reassuring to see that there is a correspondence between VLBI and Gaia results (see also \cite{ortizgaia}. But when trying to understand the statistical properties, we have found that a Gaussian error distribution can represent the distribution quite well (Fig.~2), provided we allow for an offset of the Gaia parallax and take the excess noise of the Gaia results into account \cite{lindegren18}. The reason for the large residuals in the Gaia parallaxes is understood to originate from the nature of the AGB stars; their surface brightness distribution and colour is variable, resulting in a shift of centroid that adds significantly to the positions that Gaia measures \cite{chiavassa18}. Clearly, for determining the distances and deriving quantitative properties of Mira variables and other AGB stars, VLBI astrometry still provides very valuable data.

\section{The distribution of high mass star formation}

The precise location and motion of high HMSF regions throughout the Galaxy can be measured with VLBI observations of methanol (at 6.7 and 12 GHz ) and water masers at 22 GHz. The BeSSeL (Bar and Spiral Structure Legacy) survey has produced a census of the size and rotation of the Galaxy, including the location of its spiral structure \cite{reid14}. The effort continues by measuring more sources with the new 6.7 GHz capability on the VLBA (e.g. \cite{immerevn18}). Moreover, interesting results can be expected with the new capabilities in Australia to monitor Southern targets \cite{krishnan17}. The development of the SKA will eventually allow one to survey much weaker sources \cite{riojaevn18,quirogasim}. 

\begin{figure}[t]
\centering
\includegraphics[width=.5\textwidth]{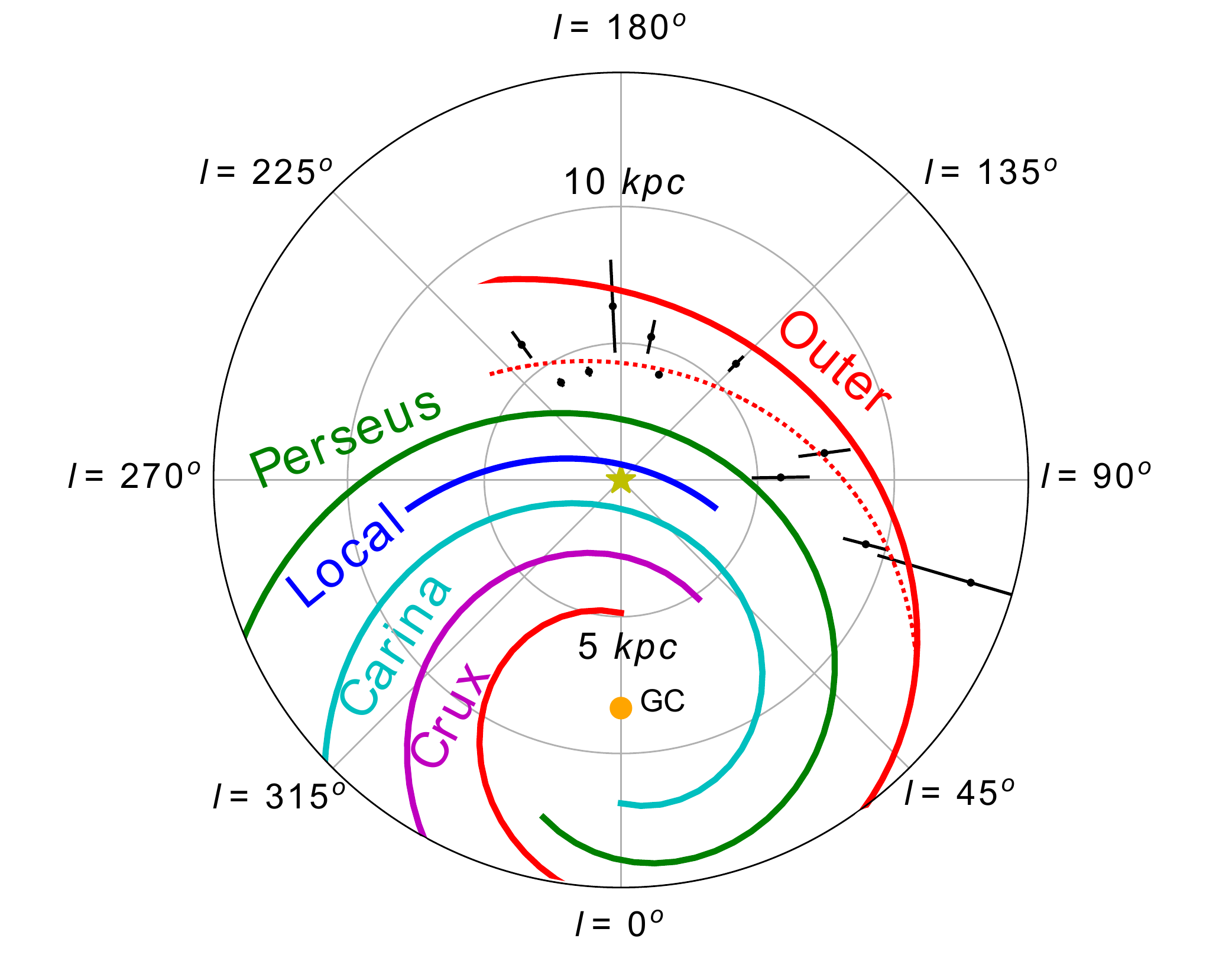}
\caption{An analysis of Outer arm star forming regions may suggest that this segment is closer than previously thought \cite{quirogas269}}
\label{fig3}
\end{figure}

The maser sites provide a unique way to outline the spiral arms of the Galaxy, as they are made of GMCs in which massive stars form. An example of a recent result is presented in Fig.~3, where the masers are used to constrain the Outer spiral arm \cite{quirogas269}. For isolated star forming regions, it may be possible to find associated members in the Gaia database, which could provide additional constraint on distances, motions and stellar properties. Although it is expected that optically identified OB associations also map out the spiral arms, Gaia will probably only be able to do that for local structures \cite{xu18}. 

\section{The distribution of evolved stars}

Maybe the most intriguing dynamical structure of the Galaxy is the bar, for which the evidence from IR photometry is convincing (e.g. \cite{babusiauxgilmore}). This structure is hardly accessible for Gaia as most of the low latitude range is blocked by interstellar extinction (Fig.~4). However, the BAaDE (Bulge Asymmetries and Dynamic Evolution) project provides a clever way to define a sample of up to $\approx$30,000 stars, selected from IR properties, which very often show show SiO maser emission from their circumstellar shells \cite{baadesjouwerman}. Indeed, a preliminary analysis has already demonstrated that a distinct population seems to make up the Mira stars associated with the bar \cite{baadetrapp}.

If it can be demonstrated that VLBI monitoring can provide proper motions (and maybe parallaxes) this would provide a unique way to determine the orbits of stars that make up the Galactic bar and bulge. VLBI astrometry of SiO masers, however, is hard because of the short coherence times for phase referencing at 43 or 86 GHz. This is particularly true for the inner Galaxy, if one tries to observe from the Northern hemisphere using the currently available networks at 7mm.  In addition, the calibrators at high frequency are sparse and weak. Finally, the special survey mode that is used for the  
BAaDE survey results a-priori maser positions that are poor by VLBI standards \cite{baadepihlstrom}. Therefore, carrying out astrometry for the BAaDE targets may in fact require new VLBI techniques \cite{iniyanevn18}.

Although optical studies cannot deliver the astrometry for stars in the bar, Gaia results are extremely important to characterise the Mira population that is present in the radio sample. We are carrying out a study of the Mira and related variables that are in the BAaDE catalogue and have Gaia counterparts. Combining with IR surveys we can derive intrinsic properties, like luminosity, IR colours and eventually variability period. These properties can be used to estimate progenitor mass and age. As stars with a wide range of mass go through the AGB phase, this will allow us to distinguish between relatively young stars and much older populations \cite{quirogagaia}.

\begin{figure}[t]
\centering
\includegraphics[width=.8\textwidth]{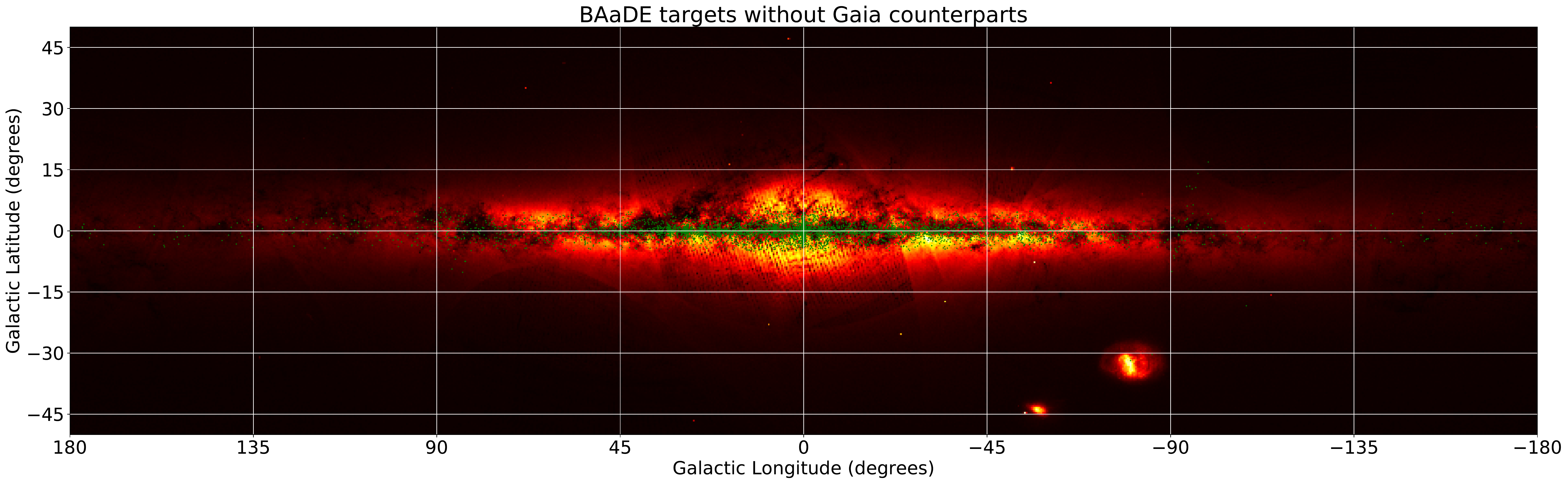}
\caption{The location of IR-selected BAaDE targets, evolved stars that likely have SiO masers, plotted over the distribution of stars in the inner Galaxy as found the Gaia database. Clearly many targets are found in the obscured, dark regions \cite{quirogagaia}.}
\label{fig4}
\end{figure}

\section{Conclusions}

We argue that VLBI astrometry will continue to deliver unique astrometry of evolved stars and obscured high mass star forming sites. This will be important for providing direct distances to the most obscured objects, which are the most extreme in terms of stellar evolution. Moreover, VLBI is not hampered by interstellar extinction and will sample the hidden parts of the Galaxy, which appear to be key in understanding its assembly.

\end{document}